\documentclass{article}

\newtheorem{theorem}{Theorem}

\newcommand{\be}{\begin{equation}}
\newcommand{\ee}{\end{equation}}
\newcommand{\bea}{\begin{eqnarray}}
\newcommand{\eea}{\end{eqnarray}}

\begin{document}

\title{\bf Resampled random processes in gravitational-wave data analysis}
\author{A. Kr\'{o}lak \thanks{Electronic Address: krolak@impan.gov.pl} \\
Institute of Mathemathics, Polish Academy of Sciences, \'{S}niadeckich 8,\\
00-950 Warsaw, Poland 
\and Massimo Tinto \thanks{Electronic Address: Massimo.Tinto@jpl.nasa.gov} \\
Jet Propulsion Laboratory, California Institute of Technology, Pasadena,\\
California 91109, U.S.A.}
\date{}
\maketitle

\begin{abstract}

The detection of continuous gravitational-wave signals requires to
account for the motion of the detector with respect to the solar
system barycenter in the data analysis.  In order to search
efficiently for such signals by means of the fast Fourier transform
the data needs to be transformed from the topocentric time to the
barycentric time by means of resampling.  The resampled data form a
non-stationary random process. In this communication we prove that
this non-stationary random process is mathematically well defined, and
show that generalizations of the fundamental results for stationary
processes, like Wiener-Khintchine theorem and Cram\`{e}r
representation, exist.

\end{abstract}

\section{Introduction}

Continuous gravitational-wave signals originating for example from
spinning neutron stars are believed to be very weak. In order to
detect them in the data of both bar and laser interferometric
gravitational-wave detectors one needs to integrate the data for many
days in order to achieve a signal-to-noise ratio sufficiently high to
guarantee their detection. Over such time scales, however, the motion
of the detector relative to the gravitational wave source introduces a
Doppler modulation in the received signal.  In order to effectively
preserve coherent integration of the signal received by the detector,
it is necessary in the data analysis to model very accurately its
motion with respect to the solar system barycenter (SSB).

The simplest model of a monochromatic gravitational-wave
signal emitted by such a source has the following form
\be
s = A\cos[\phi_0 + 2\pi f_0 t
+ \frac{2\pi}{c} f_0 {\bf n}_0\cdot{\bf r}_{\rm SSB}(t)]
\label{Eq.:sig}
\ee In Eq.\ (\ref{Eq.:sig}) $\phi_0$ is the initial phase of the
waveform, ${\bf r}_{\rm SSB}$ is the vector joining the solar system
barycenter (SSB) with the detector, ${\bf n}_0$ is the constant unit
vector in the direction from the SSB to the neutron star. We assume
that the gravitational wave form is monochromatic with frequency $f_0$
which we define as the instantaneous frequency evaluated at the SSB at
$t=0$. In general the frequency of the signal may change as a
consequence of the spindown of the rotating neutron star. However,
this does not introduce new qualitative features into the mathematical
model we discuss here.  

The above signal cannot be detected by means of the
fast Fourier transform algorithm (FFT) because the unknown frequency
$f_o$ to be estimated is contained in the highly non-linear function
of time depending on the position of the detector with respect to the
SSB. Consequently the following technique is
proposed$^{\cite{BCCS98}}$. Introduce a new time $t_B$ called
barycentric time which is related to the topocentric time $t$ by \be
t_B = t + {\bf n}_0\cdot{\bf r}_{\rm SSB}(t)/c.
\label{TB}
\ee
Then the signal given above takes the form
\be
s_B = A\cos(\phi_0 + 2\pi f_o t_B).
\label{Eq.:sigB}
\ee The signal $s_B$ is monochromatic and can be searched for by means
of FFT.  However, as we shall show in the following section, the
transformation given by Eq. \ref{TB} makes the noise in the data a
non-stationary random process.  The main result of this communication,
proven in the next section, will be to show that this non-sationary
random process is mathematically well defined and generalizations of
the fundamental results for stationary processes like
Wiener-Khintchine theorem and cram\`{e}r representation exist. We also
show that, if the spectrum of the unresampled noise is white, the
resampled process remains stationary.

Another case when the resampling of the data is applied are the
so-called "accelarated searches" in pulsar data
analysis$^{\cite{JK91,L99}}$.  These are searches for pulsars in
binary systems where the observation time is much shorter than the
period of the orbit so that we can expand the phase of the signal in
Taylor series keeping only the second order terms.  This results in
the following signal \be s_a = A\cos[\phi_0 + 2\pi (f_o t + p_o t^2)].
\label{Eq.:siga}
\ee This signal can be detected by matched filtering i.e. by
multiplication of the data by $exp(-2\pi p t^2)$ for a grid of
prameters $p$ followed by FFT. Nevertheless pulsar astronomers apply
resampling also in this case.  One defines a new time \be t_a = t + a
t^2, \ee where $a = p_o/f_o$. Then the signal $s_a$ becomes
monochromatic: \be s_a = A\cos[\phi_o + 2\pi f_o t_a] \ee and can be
searched for by means of FFT. Of course we do not know the frequency
$f_o$ and parameter $p_o$, and therefore have to resample the data
over an appropriate grid on the acceleration parameter $a$.

\section{Resampled random processes}

Let $x(t)$ be a continuous parameter real-valued random process. The process $%
x(t)$ is said to be {\em completely stationary}$^{\cite{PW1993}}$
(sometimes referred to as {\em strongly stationary} or {\em strictly
stationary) }if for all $n\geq 1$, for any $t_{1,}t_{2,...,}t_{n}$ contained
in the index set, and for any $\tau $ such that $t_{1}+\tau ,t_{2}+\tau
,...,t_{n}+\tau $ are also contained in the index set, the joint cumulative
probability distribution function(cpdf) of $x(t_{1}),x(t_{2}),...,x(t_{n})$
is the same as that of $x(t_{1}+\tau ),x(t_{2}+\tau ),...,x(t_{n}+\tau ).$
In other words, the probabilistic structure of a completely stationary
process is invariant under a shift of time. The process $x(t)$ is said to be 
{\em second-order stationary}$^{\cite{PW1993}}$ (sometimes
referred to as {\em weakly stationary} or {\em covariance stationary) }if
for all $n\geq 1$, for any $t_{1,}t_{2,...,}t_{n}$contained in the index
set, and for any $\tau $ such that $t_{1}+\tau ,t_{2}+\tau ,...,t_{n}+\tau $
are also contained in the index set, the joint moments of orders 1 and 2 of $%
x(t_{1}),x(t_{2}),...,x(t_{n})$ exist, are finite and are equal to
corresponding joint moments of $x(t_{1}+\tau ),x(t_{2}+\tau
),...,x(t_{n}+\tau ).$ A weakly stationary Gaussian random process is also
completely stationary because a Gaussian process is completely determined by
its 1st and 2nd moment. For stationary random processes we have the
following fundamental result.$^{\cite{P1981}}$

\begin{theorem}[The Wiener-Khintchine Theorem]
\label{Th_WK}

A necessary and sufficient condition for $\rho (\tau )$ to be the
autocorrelation function of some stochastically continuous (i.e. continuous
in the mean square sense) stationary process, $x(t)$, is that there exists a
function, $F(\omega )$, having the properties of a distribution function on $%
(-\infty ,\infty )$, (i.e. $F(-\infty )=0,F(\infty )=1,$and $F(\omega )$
non-decreasing), such that, for all $\tau $,$\rho (\tau )$ may be expressed
in the form, 
\begin{equation}
\rho (\tau )=\int_{-\infty }^{\infty }e^{i\omega \tau }dF(\omega ).
\end{equation}
\end{theorem}

The necessary part of the above Theorem follows from a general theorem
due to Bochner that any positive semi-definite function which is
continuous everywhere must have the representation of the above form.
In the case of a purely continuous spectrum we have $dF(\omega ) =
S(\omega )d\omega$, where $S(\omega )$ is the (normalized) spectral
density function. Thus the Wiener-Khintchine theorem asserts that a
well defined spectral density exists.

Another important result is the existence of the spectral decomposition 
of the stationary random process itself.

\begin{theorem}[The Cram\`{e}r representation]
\label{Th_Cr}

Let $x(t)$,$-\infty <t<\infty ,$be a zero-mean stochastically continuous
stationary process. Then there exists an orthogonal process, $Z(\omega ),$
such that for all $t$, $x(t)$ may be written in the form, 
\begin{equation}
x(t)=\int_{-\infty }^{\infty }e^{i\omega t}dZ(\omega ),
\end{equation}
the integral being defined in the mean-square sense. The process $Z(\omega )$
has the following properties;

(i) $\ E[dZ(\omega )]=0,$ for all $\omega $

(ii) $E[|dZ(\omega )|^{2}]=dH(\omega ),$ for all $\omega ,$

where $H(\omega )$ is the (non-normalized) integrated spectrum of $x(t)$,

(iii) for any two distinct frequencies, $\omega $ , $\omega ^{\prime }$, $%
(\omega \neq \omega ^{\prime }),$%
\begin{equation}
cov[dZ(\omega ),dZ(\omega ^{\prime })]=E[dZ^{\ast }(\omega )dZ(\omega ^{\prime })]=0.
\end{equation}
\end{theorem}
The above result says that any stationary random process can be decomposed into 
a sum of sine and cosine functions with uncorrelated coefficients.

Let $x(t)$ be a random process and let $t_{r}=t+k(t;\theta )$ be a
smooth one-to-one function both of the index $t$ and the parameter set
$\theta $. A
as the random process $x(t)$ taken at time $t_{r}$
i.e. $y(t_{r})=x(t).$ In the following for simplicity we assume that
the process $x(t)$ is zero-mean. It immediately follows that the
resampled process $y(t_{r})$ is also zero mean. Suppose that the
original process is stationary. Let us first convince ourselves that
the resampled process is, in general, {\em non-stationary. } Let $%
C(t_{r}^{^{\prime }},t_{r}):=E[y(t_{r}^{^{\prime }})y(t_{r})]$ be the
autocovariance function of the resampled process. By definition of the
resampled process we have that $C(t_{r}^{^{\prime
}},t_{r}):=E[x(t^{^{\prime }})x(t)]$ and by stationarity of $x(t)$ we
have $C(t_{r}^{^{\prime }},t_{r})=R(t^{^{\prime }}-t).$ By implicit
function theorem we have that there exists a smooth function
$t=t_{r}+g(t_{r};\theta ).$ In order that the resampled process be
stationary the function $R$ must depend only on the difference
$(t_{r}^{^{\prime }}-t_{r})$ This is the case if and only if $t$ is a
linear function of $t_{r}$ i.e. $t=t_{r}+a(\theta )t_{r}+b(\theta ).$
Thus when the resampling transformation is non-linear the resulting
resampled process is non-stationary.

In the linear resampling case the Fourier transform $\widetilde{Y}(\omega
_{r})$ of the resampled process at frequency $\omega _{r}$ is related to the
Fourier transform $\widetilde{X}(\omega )$ of the original process at
frequency $\omega $ by the following formula: 
\begin{equation}
\widetilde{Y}(\omega _{r})=\frac{\exp i\omega b}{1+a}\widetilde{X}(\omega ),
\end{equation}
where $\omega =\frac{\omega _{r}}{1+a}$

Let us consider the covariance function $C(t_{r}^{^{\prime }},t_{r})$, of
the resampled process $y(t_{r}).$ It can be written as

\begin{equation}
C(t_{r}^{^{\prime }},t_{r})=\int_{-\infty }^{\infty }\phi _{t_{r}^{^{\prime
}}}^{\ast }(\omega )\phi _{t_{r}}(\omega )dH(\omega ),
\end{equation}
where we have introduced a set functions 
\begin{equation}
\phi _{t_{r}}(\omega )=\exp [i\omega (t_{r}+g(t_{r}))]
\end{equation}
We have the following important result.

\begin{theorem}[General orthogonal expansions]
Let $y(t_{r})$ be a continuous parameter zero mean process (not necessarily
stationary) with covariance function $C(t_{r},s_{r})=E[y(t_{r})y(s_{r})]$.
If there exists a family of functions, \{$\phi _{t_{r}}(\omega )\}$, defined
on the real line, and indexed by suffix $t_{r}$, and a measure, $\mu (\omega
)$, on the real line such that for each $t$, $\phi _{t_{r}}(\omega )$ is
quadratically integrable with respect to the measure $\mu $, i.e. 
\begin{equation}
\int_{-\infty }^{\infty }\left| \phi _{t_{r}}(\omega )\right| ^{2}d\mu
(\omega )<\infty 
\end{equation}
and for all $t_{r}$, $s_{r},$ $C(t_{r},s_{r})$ admits a representation of
the form, 
\begin{equation}
C(t_{r}^{^{\prime }},t_{r})=\int_{-\infty }^{\infty }\phi _{t_{r}^{^{\prime
}}}^{\ast }(\omega )\phi _{t_{r}}(\omega )dH(\omega )  \label{gWK}
\end{equation}
then the process $y(t_{r})$ admits a representation of the form 
\begin{equation}
y(t_{r})=\int_{-\infty }^{\infty }\phi _{t_{r}}(\omega )dZ(\omega ),
\label{gCR}
\end{equation}
where $Z(\omega )$ is an orthogonal process with 
\begin{equation}
E[|dZ(\omega )|^{2}]=d\mu (\omega ).
\end{equation}
\end{theorem}

Conversely if $y(t_{r})$ admits a representation of the form with an
orthogonal process satisfying , then $C(t_{r}^{^{\prime }},t_{r})$ admits a
representation of the form .

The formula (\ref{gWK}) is called{\em \ generalized Wiener-Khintchine}
relation and formula (\ref{gCR}) is called {\em generalized Cram\`{e}r
representation} of the random process. In our case the generalized Cram\`{e}%
r representaion reads
\begin{equation}
y(t_{r})=\int_{-\infty }^{\infty }\exp [i\omega (t_{r}+g(t_{r}))]dZ(\omega ),
\end{equation}
\ This representation also clearly shows that the resampled process is in
general non-stationary because the choice of basis function 
\begin{equation}
\phi _{t_{r}}(\omega )=\exp (i\omega t_{r})
\end{equation}
is not in general possible. The generalized Cram\`{e}r representation is also
immediate from the Cram\`{e}r representation for the original stationary
process and its transformation to resampled time. It also follows that the
measure $\mu $ coincides with the integrated spectrum $H(\omega )$ of the
original stationary process. However $H(\omega )$ cannot be interepreted as
the spectrum of the resampled process. Indeed for the resampled process
which is non-stationary the concept of spectrum is not mathematically well
defined.

The general orthogonal expansion theorem has already been used by M. B.
Priestley \cite{P1981} to develop the theory of so called {\em evolutionary
spectra}. This theory describes a very important class of non-stationary
processes often occurring in practice for which the amplitude of the Fourier
transform slowly changes with time.

In the case of a continuous spectrum we have $dH\left( \omega \right) =S\left(
\omega \right) d\omega$. Then one can write the properties (ii) and (iii) of the
Cram\`{e}r representation theorem as 
\begin{equation}
E[\widetilde{X}(\omega ^{^{\prime }})^*\widetilde{X}(\omega )]=S\left( \omega
\right) \delta \left( \omega ^{^{\prime }}-\omega \right) 
\end{equation}
where $\delta $ is the Dirac function. In the continuous case it is
instructive to calculate the correlation function for the Fourier frequency
components of the resampled (non-stationary) process.
Using Eq.(\ref{gCR}) the Fourier transform $\widetilde{Y}(\omega)$ of the resampled process 
$y(t_r)$ can be written as
\be
\widetilde{Y}(\omega) = \int_{-\infty }^{\infty } Q(\omega_1,\omega) dZ(\omega_1 ),
\ee
where the kernel $Q$ is given by
\be
Q(\omega_1,\omega) = \int_{-\infty }^{\infty } \phi _{t_{r}}(\omega_1 ) 
                    \exp(-i\omega t_r) dt_r.
\ee
The correlation between two Fourier components of the resampled process
takes the form
\bea
E[\widetilde{Y}(\omega ^{^{\prime }})^*\widetilde{Y}(\omega )] &=&
\int_{-\infty }^{\infty } \int_{-\infty }^{\infty } Q(\omega_1,\omega^{\prime })^* Q(\omega_2,\omega) 
E[dZ(\omega_1 )^* dZ(\omega_2 )] \nonumber \\
&=& \int_{-\infty }^{\infty } Q(\omega_1,\omega^{\prime} )^* Q(\omega_1,\omega) S(\omega_1) d\omega_1
\eea
Thus we see that for a resampled random process the Fourier components at different frequencies are
correlated, This is another manifestation of the non-stationarity of the process.
Let us now consider the example of white noise for which the spectral density is independent
of frequency $\omega$. It is straightforward to show that
\be
\int_{-\infty }^{\infty } Q(\omega_1,\omega^{\prime })^* Q(\omega_1,\omega) d\omega_1 =
\delta(\omega^{\prime } - \omega)
\ee
Thus for the case of white noise we have that
\be
E[\widetilde{Y}(\omega ^{^{\prime }})^*\widetilde{Y}(\omega )] = S\left( \omega
\right) \delta \left( \omega ^{^{\prime }}-\omega \right) 
\ee
and consequently in this case the noise remains stationary after resampling.
It is also easy to see that the Fourier components at different frequencies
will be uncorrelated if the spectral density is constant over the
bandwidth of the kernel $Q$. It is possible that the last assumption will be
fulfilled in the case of search of the data from gravitational-wave detectors
from continuous sources$^{\cite{BCCS98a}}$ but is not guaranteed. 

\section*{Acknowledgments}

We would like to thank Prof. Bernard F. Schutz for illuminating
discussions concerning nonstationarity of the resampled noise. This
work was supported in part by the KBN Grant No.\ 2 P03B 094 17.
This research was performed at the Jet Propulsion Laboratory,
California Institute of Technology, under contract with the National 
Aeronautics and Space Administration (M.T.).

\end{document}